\documentclass[twocolumn,prl,superscriptaddress,preprintnumbers]{revtex4}
\pdfoutput=1
\usepackage{amsmath,amssymb,graphicx,amsfonts}
\usepackage{epsf,verbatim}
\usepackage{hyperref}
\usepackage{color}
\usepackage{natbib}

\newcommand{\simgt}{\mathrel{\lower2.5pt\vbox{\lineskip=0pt\baselineskip=0pt
           \hbox{$>$}\hbox{$\sim$}}}}
\newcommand{\simlt}{\mathrel{\lower2.5pt\vbox{\lineskip=0pt\baselineskip=0pt
           \hbox{$<$}\hbox{$\sim$}}}}

\newcommand{\squishlist}{
 \begin{list}{$\bullet$}
  { \setlength{\itemsep}{0pt}
     \setlength{\parsep}{3pt}
     \setlength{\topsep}{3pt}
     \setlength{\partopsep}{0pt}
     \setlength{\leftmargin}{1.5em}
     \setlength{\labelwidth}{1em}
     \setlength{\labelsep}{0.5em} } }

\newcommand{\squishlisttwo}{
 \begin{list}{$\bullet$}
  { \setlength{\itemsep}{0pt}
     \setlength{\parsep}{0pt}
    \setlength{\topsep}{0pt}
    \setlength{\partopsep}{0pt}
    \setlength{\leftmargin}{2em}
    \setlength{\labelwidth}{1.5em}
    \setlength{\labelsep}{0.5em} } }

\newcommand{\squishend}{
  \end{list}  }

\begin{document}

\title{Implications of Higgs Searches on the Four Generation Standard Model}

%\preprint{YITP-SB-01-12}

\author{Eric Kuflik} \thanks{ekuflik@gmail.com}
\affiliation{Raymond and Beverly Sackler School of Physics and
  Astronomy, \\Tel-Aviv University, Tel-Aviv 69978, Israel}

\author{Yosef Nir}
\thanks{yosef.nir@weizmann.ac.il}
\affiliation{Department of Particle Physics and Astrophysics, Weizmann
  Institute of Science,  Rehovot 76000, Israel}

\author{Tomer Volansky} \thanks{tomerv@post.tau.ac.il}
\affiliation{Raymond and Beverly Sackler School of Physics and
  Astronomy, \\Tel-Aviv University, Tel-Aviv 69978, Israel}

\begin{abstract}
  Within the four generation Standard Model, the Higgs couplings
  to gluons and to photons deviate in a significant way from the
  predictions of the three generation Standard Model.  As a consequence,
  large departures in several Higgs production and decay channels are
  expected.  Recent Higgs search results, presented by ATLAS, CMS and
  CDF, hint on the existence of a Higgs boson with a mass around 125
  GeV.  Using these results and assuming such a Higgs boson, we
  derive exclusion limits on the four generation Standard Model.
  For $m_H=125$ GeV, the model is excluded at 99.9\%
  confidence level. For $124~{\rm GeV} \le m_H \le 127$ GeV, an exclusion limit
  above 95\% confidence level is found.
\end{abstract}

\maketitle

 \setcounter{equation}{0} \setcounter{footnote}{0}

%%%%%%%%%%%%%%%%%%%%%%%%%%%%%%%%%%%%%%%%%%%%%%%%%%%
\section{INTRODUCTION}
\label{sec:intro}
%%%%%%%%%%%%%%%%%%%%%%%%%%%%%%%%%%%%%%%%%%%%%%%%%%%
%
The intriguing possibility of a four generation Standard Model (SM4) has
been studied intensively (see {\it e.g.}~\cite{Holdom:2009rf}
and references therein).  Constraints on this scenario
arise directly, via the search for production of fourth
generation quarks and leptons at colliders~\cite{Aad:2012bb,Aad:2012us},
and indirectly, through their effect on the oblique electroweak parameters
\cite{Peskin:1991sw,Kribs:2007nz,Hashimoto:2010at} and on the Higgs
boson production and decay partial
widths~\cite{Guo:2011ab,Chen:2012wz}.  In the context of the latter set
of observables, it has long been
realized that the presence of a fourth generation drastically changes
the Higgs branching fractions.  In particular, the couplings to gluons
and to photons are induced at the loop level and are therefore
susceptible to the presence of (respectively, colored and electromagnetically
charged) heavy new particles.  As a consequence, precise
measurements of the Higgs production rate and branching ratios can
strongly constrain the existence of a fourth generation.

Recently, the ATLAS, CMS, CDF and D0 experiments have reported new
results~\cite{ATLAS:2012ac,ATLAS:2012ad,ATLAS:2012ae,Chatrchyan:2012tw,Chatrchyan:2012dg,Chatrchyan:2012tx,D0-HIG}
which hint on the existence of a light Higgs boson with a mass of
order 125~GeV.  Several Higgs decay channels have been probed,
including the $\gamma\gamma$, $ZZ^*$ and $WW^*$ channels dominated by
the gluon fusion production mode, $b\bar b$ in the associated
production mode, and diphoton in association with two jets channel which has
a large vector boson fusion (VBF) production component.  These results
favor a somewhat large rate in all but the $WW^*$ channel, where no significant
excess is found. In this letter we analyze these results under the
assumption that a Higgs signal has been observed. As we show below,
under such an assumption, the SM4 is excluded.

Three ingredients are important in making such an exclusion possible.
First, the fourth generation top and bottom quarks would {\em enhance}
the gluon fusion production rate of a light Higgs boson by a factor of
$\mathcal{O}(10)$~\cite{Georgi:1977gs}.  Second, the partial decay
width to diphotons can be {\em suppressed} by as much as a factor of
$\mathcal{O}(100)$~\cite{Denner:2011vt}.  Third, partial decay widths
to final states which are dominated by tree-level amplitudes, such as
$b\bar b$ and $ZZ^*$, receive smaller corrections to the Standard Model
(SM) prediction.  The net result is therefore a significant enhancement
in all gluon fusion produced channels, with the exception of the
diphoton channel which is significantly suppressed.

The data discussed above favors enhanced rates, but not as high
as predicted in the SM4.  This situation has led the CMS
collaboration to rule out the SM4 for $m_h > 120$ GeV at 95\%
CL and for $m_h > 125$ GeV at 99\% CL \cite{CMS-PAS-HIG-12-008}.  The
CMS analysis, however, assumes that the fourth generation neutrino is
heavy enough so that the additional invisible Higgs decay mode is
forbidden, $m_N\simgt m_h/2$.  The inclusion of such a channel dilutes
all branching fractions uniformly and hence significantly weakens
the CMS exclusion limit.

In this note we relax the assumption on the mass of the fourth
neutrino. Yet, we obtain significantly stronger exclusion limits and conclude
that, in the presence of a light Higgs boson, the four
generation Standard Model is excluded \cite{Tomertalk}.

%%%%%%%%%%%%%%%%%%%%%%%%%%%%%%%%%%%%%%%%%%%%%%%%%%%
\section{The SM4 Rates}\label{sec:rates}
In the SM, the gluon fusion amplitude is dominated by the
top-induced one-loop contribution. The SM4 introduces two new heavy
quarks into the loop, for which the leading-order (LO) contribution is
approximately independent of the actual masses. Consequently, the
gluon fusion rate is enhanced by a factor of 9 at LO.

The fourth generation top and bottom modify also the LO contributions
to the Higgs partial widths to digluons and diphotons. The latter is also
affected by the fourth generation charged lepton.
Similarly to the gluon fusion production cross-section, the
$h\rightarrow gg $ width is increased by a factor $~9$.  On the other
hand, $h\rightarrow \gamma\gamma $, which is dominated by $W$-boson
loop, is suppressed as the additional fermions interfere destructively
with the $W$-boson contribution. At LO this amounts to decreasing the
diphoton width by a factor of about $5$ relative to the
SM; it is also mostly independent of fermion masses. Finally, the
other leading partial widths, which are all allowed at tree-level,
remain unchanged at LO.

At next-to-leading-order (NLO), the large Yukawa-couplings for heavy
fermions can contribute significantly to all widths. Complete NLO
widths have been calculated by Denner {\it et al}~\cite{Denner:2011vt},
and partially implemented in  HDECAY~\cite{Djouadi:1997yw} and
Prophecy4f~\cite{Bredenstein:2007ec}. For very heavy fermion masses,
up to the perturbative limit, the corrections to the decay rates to fermions
and heavy gauge bosons  can be as large as a factor of 2, and tend to increase
the widths to fermions and decrease the widths to $WW^{\star}$ and $ZZ^{\star}$.
The NLO corrections to $h\rightarrow gg $ are less significant.

The LO value of the $h\rightarrow \gamma \gamma$ width is already
accidentally small due to the destructive interference between the
$W$-boson and fermion loops. As a consequence, the NLO corrections are
relatively large, and the two-loop matrix-elements can lead to another
significant {\emph{cancelation}} in the amplitude. For instance, for
$m_h = 130$ GeV and for fermion masses given in the ``extreme
scenario'' of Ref.~\cite{Denner:2011vt}, the cancelation between the LO and
NLO correction is 90.8\%.

HDECAY approximates the relative NLO corrections of $h\rightarrow
\gamma \gamma$ to about 1\% accuracy. However due to the very large cancelation,
this  may result in an $\mathcal{O}(1)$ inaccuracy in the actual
width at NLO. Additional sources of theoretical error/uncertainty
arise in the NNLO corrections which may be as large as $100\%$. As
discussed below, for light fourth generation fermion masses, where
the Higgs constraints are
the {\it weakest}, these uncertainties are expected to be low. For all cases, we
calculate the widths at $m_h =$ 120, 125, and 130 GeV and
interpolate the widths for intermediate Higgs masses.

%%%%%%%%%%%%%%%%%%%%%%%%%%%%%%%%%%%%%%%%%%%%%%%%%%%
\section{HIGGS SEARCHES AT COLLIDERS} \label{sec:LHC}
Recently, the CMS, ATLAS, CDF and D0 experiments have reported results
of Higgs searches in various
channels~\cite{ATLAS:2012ac,ATLAS:2012ad,ATLAS:2012ae,Chatrchyan:2012tw,Chatrchyan:2012dg,Chatrchyan:2012tx,
 D0-HIG}. Three of the experiments report an excess
of events which hint of the existence of a Higgs boson around 125 GeV.
The excess is mostly apparent in four channels: inclusive diphoton,
diphoton in association with two jets, fully leptonic $ZZ^*$, and
associated production of Higgs decaying to $b\bar b$.  On the other hand,
there is no apparent signal in Higgs decay to $WW^*$.

The gluon fusion production (which is expected to be the
dominant source of Higgs bosons at the LHC) and the diphoton decay
are particularly sensitive to the presence of additional
sequential quarks and leptons. Hence measurements of these rates provide an
excellent opportunity to revisit the limits on the SM4.  The
excess observed indicates a cross-section that is somewhat larger than the
SM prediction.  While this cannot (and at present should
not) be taken as a hint for new physics, it can be used to put strong
constraints on the SM4.

In order to efficiently constrain the SM4, it is crucial to consider each
Higgs search channel separately. In Ref.~\cite{Carmi:2012yp}, a combination
of the ATLAS and CMS results is presented for the five channels mentioned above.
For each channel $i$, the best fit value for the signal strength, $\hat \mu_i$, is
derived (or, when possible, directly taken from the reported results)
as a function of the Higgs mass, by maximizing the corresponding
likelihood functions.  $\hat\mu_i$ can then be compared with the
corresponding SM4 value $R_i$.  While the combination has not been
presented by the collaborations and should be used with caution, the
results are expected to be conservative. For $m_h$ between 120 GeV and 130 GeV,
all $\hat\mu_i$ and the corresponding standard deviations $\sigma_i$ are given
by the ATLAS and CMS collaborations,
with the exception of the diphoton plus dijet channel, in which $\hat\mu$ and $\sigma$
are only provided for $m_h = 125$ GeV by CMS; ATLAS did not report results for the diphoton plus dijet channel. The full mass range was calculated in \cite{Carmi:2012yp}.
We do not use the $\gamma\gamma jj$ results of the more recent multivariate analysis \cite{CMSMVA},
since the relative efficiencies for different production modes are not stated.

%%%%%%%%%%%%%%%%%%%%%%%%%%%%%%%%%%%%%%%%%%%%%%%%%%%
\section{RESULTS} \label{sec:results}
\begin{figure*}[t]
\begin{center}
  \includegraphics[width=.45\textwidth]{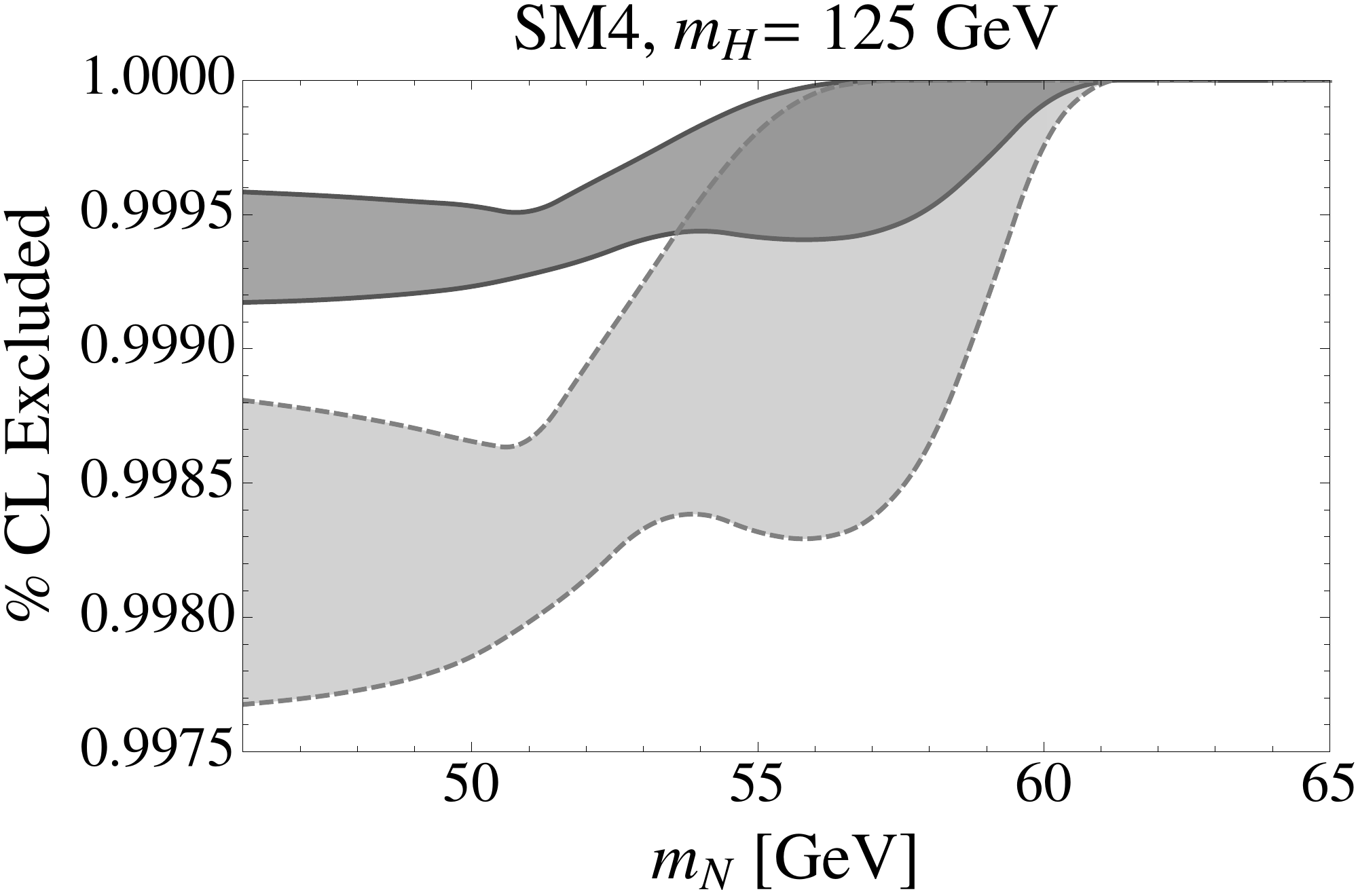}\qquad
  \qquad
  \includegraphics[width=.45\textwidth]{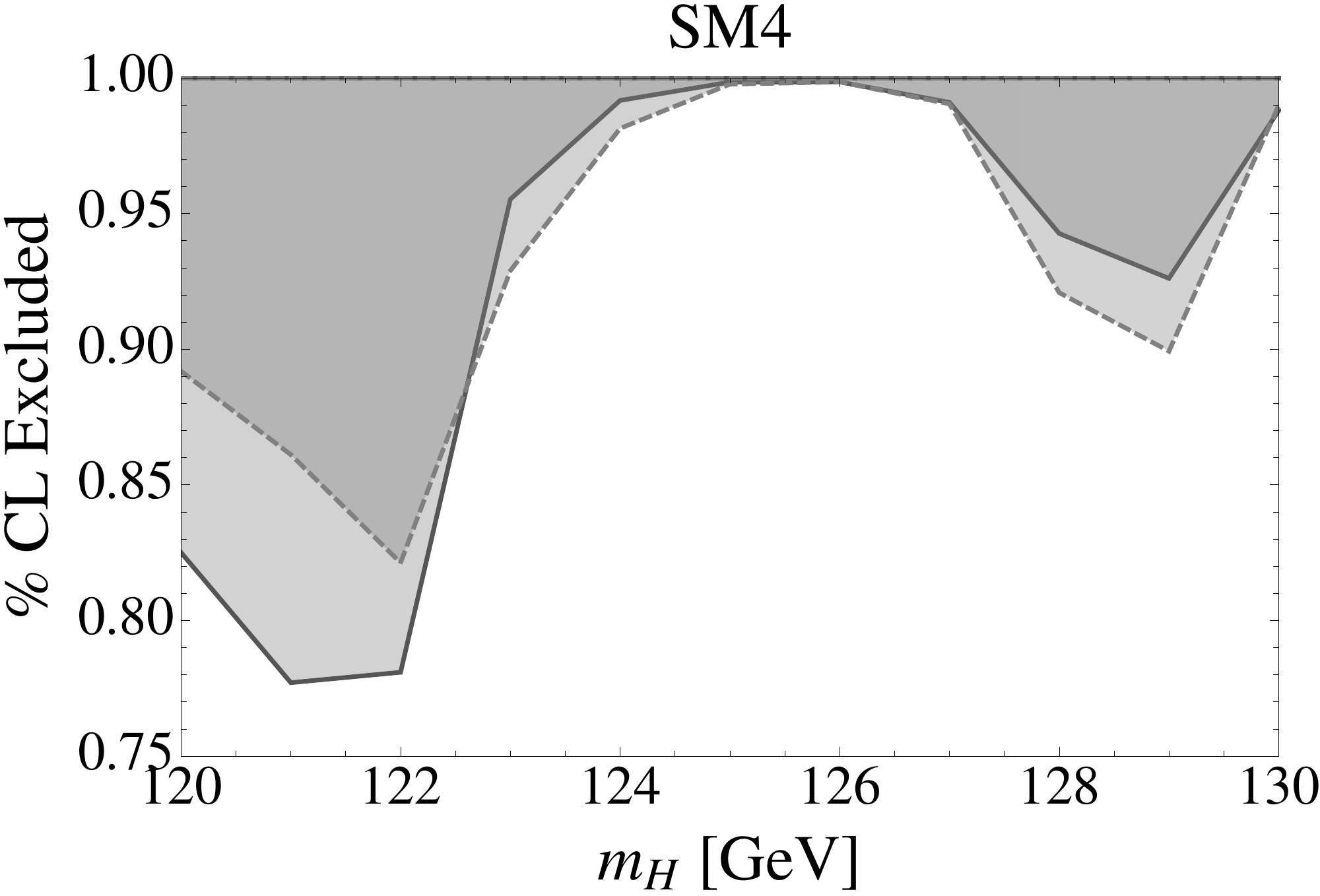}
\caption{Constraints on the SM4 derived from scanning over the fourth generation
fermion masses as described
in the text. The darker region within the solid borders (light region within the
dashed borders) shows the level of exclusion with (without) the $\gamma \gamma jj$
mode.  {\bf Left}:  the exclusion limit as a function of the neutrino mass
(the other fourth generation fermion masses are scanned) for fixed $m_h=125$ GeV.
{\bf Right}: the exclusion limit as a
 function of $m_h$, with all fourth generation masses varied.}
%For the $\gamma \gamma jj$ channels, the values calculated in \cite{Carmi:2012yp} are used.}
  \label{fig:res}
\end{center}
\end{figure*}

In Fig.~\ref{fig:res}, we show the exclusion limits on the SM4,
using the LHC and Tevatron Higgs measurements discussed above.
The shaded regions show the results of a scan over SM4 spectra.
All masses are required to be below the ``extreme scenario'' of
Ref.~\cite{Denner:2011vt}  where perturbitivity reaches its limit.
Additionally, our scan includes only sets of parameters that
are within the 95\% CL ellipse of the S and T oblique
parameters~\cite{Kribs:2007nz,Hashimoto:2010at,Baak:2011ze,Denner:2011vt}.

The constraints are made by minimizing the $\chi^2$,
\begin{equation}
  \chi^2 = \sum_{\rm channels}\frac{\left(R_{\rm i}-\hat{\mu}_{\rm
        i}\right)^2} {\sigma^2_{\rm i}}.
\end{equation}
The sum runs over the five measured channels: inclusive diphoton,
diphoton in association with two jets, $ZZ^*$, $WW^*$, and $b\bar b$
in association with a vector-boson. We assume that individual likelihoods
follow a gaussian distribution, when calculating the $\chi^2$ cumulative
distribution functions.

On the left box of Fig.~\ref{fig:res}, we show the exclusion limit as
a function of $m_N$, the fourth generation neutrino mass, for fixed $m_h = 125$ GeV.
The darker region shows the confidence level exclusion when including
the $\gamma\gamma jj$ mode, while the lighter region shows the exclusion when omitting this mode.
Since there are large systematic uncertainties in the gluon-fusion contribution to the dijet mode,
we show constraints with this mode separately. When including (omitting)
the $\gamma\gamma jj$ mode, the SM4 with $m_h \ge 125$ GeV is excluded at above
99.9\%  (99.7\%) CL, for all values of the fourth generation neutrino mass.

On the right box of Fig.~\ref{fig:res}, we show the exclusion limit as
a function of the Higgs mass. Again we show the exclusions with and without the
$\gamma\gamma jj$ mode. Since the CMS experiment does not provide the values of
$\hat{\mu}_{\gamma \gamma jj}$ for $m_h = 120-130$
GeV, we use the analysis of \cite{Carmi:2012yp} where the best fit rates and
errors are calculated. For $m_h > 123$ GeV, the SM4 is excluded above $90\%$ CL. For
$124 ~{\rm GeV} < m_h < 127~{\rm GeV} $, the SM4 is excluded above 95\% CL.

The numerical scan shows robust exclusion over all of the parameter space.
As mentioned above, care should be taken with these numerical codes as they only
approximately calculate $\Gamma(h\rightarrow \gamma \gamma) $ at NLO, and NNLO
corrections may be large for the heavier masses scan.   Nonetheless,  given these
results,  the constraints are expected to remain strong even if exact calculations
could be performed. Indeed, we note that the weakest constraints are obtained
when the fourth generation masses are lightest.  This is intuitive, since smaller
Yukawa couplings imply smaller corrections, and consequently a smaller
cancelation in $h \rightarrow \gamma \gamma $ width.   However, in precisely this
region, the uncertainties in using the numerical code to
calculate the width and the unknown NNLO corrections are both expected to be small.
Thus, we do not expect these corrections to significantly alter the results  obtained
from the scan.

%%%%%%%%%%%%%%%%%%%%%%%%%%%%%%%%%%%%%%%%%%%%%%%%%%%
\section{DISCUSSION} \label{sec:summary}
The Higgs boson is yet to be discovered.  Nonetheless, evidence from
three independent experiments, ATLAS, CMS and CDF, hint to its
existence and pointing to a mass of about 125 GeV. Under the assumption
that these measurements are not the result of a statistical
fluctuation, stringent constraints on the low energy effective
couplings of the Higgs boson to heavy quarks and vector bosons can be
placed~\cite{Carmi:2012yp,Azatov:2012bz,Espinosa:2012ir,Giardino:2012ww,Ellis:2012rx}.

A fourth generation would affect strongly the Higgs
effective couplings to gluons and photons and consequently the
corresponding Higgs partial decay widths.  Concretely, the gluon fusion rate is enhanced by a factor $\sim9$ while the diphoton decay rate is suppressed
by a factor $\sim100$.  Consequently, several decay channels, such as $ZZ^*$ and $WW^*$, which are dominantly produced via gluon fusion, are predicted to be
enhanced, contradicting current measurements.  It is possible to ameliorate the tension in these channels by allowing the fourth generation neutrino to be light,
thereby uniformly suppressing all branching fractions.  However, the already suppressed diphoton channel is then far below its measured value.

The reasoning above allows one to strongly exclude the four generation
Standard Model.   For a Higgs mass of 125 GeV, we find it to be excluded
at the 99.9\% CL.

\begin{center}

{\bf Acknowledgements}
\end{center}
The authors thank Gideon Bella, Ansgar Denner, Adam Falkowski, Zoltan
Ligeti and Michael Spira for useful discussions. The work of EK and TV
is supported in part by a grant from the Israel Science Foundation.
The work of TV is further supported in part by the US-Israel
Binational Science Foundation and the EU-FP7 Marie Curie, CIG
fellowship. YN is the Amos de-Shalit chair of theoretical physics and
is supported by the Israel Science Foundation and by the
German-Israeli foundation for scientific research and development (GIF).

%%%%%%%%%%%%%%%%%%%%%%%%%%%

%
%\bibliography{SM4}{}
%\bibliographystyle{h-physrev}

\end{document}